\documentclass[conference]{IEEEtran}
\IEEEoverridecommandlockouts
\usepackage{cite}
\usepackage{amsmath,amssymb,amsfonts}
\usepackage{booktabs}
\usepackage{amsmath}
\usepackage{algorithmic}
\usepackage{graphicx}
\usepackage{textcomp}
\usepackage{xcolor}
\usepackage{hyperref}
\def\BibTeX{{\rm B\kern-.05em{\sc i\kern-.025em b}\kern-.08em
    T\kern-.1667em\lower.7ex\hbox{E}\kern-.125emX}}
\begin{document}

\title{HPC-Enabled Video-based Coastal Wave Parameter Estimation Using V-JEPA and Deep Spatiotemporal Learning\\

\thanks{ 979-8-3195-0703-7/26/\$31.00~\textcopyright~2026 IEEE }
}

\author{%
  \IEEEauthorblockN{Abubakar Hamisu Kamagata}
  \IEEEauthorblockA{\textit{Department of Computer Science}\\
  \textit{Namibia University of Science and Technology}\\
  Windhoek, Namibia\\
  kamagata012@gmail.com}
  
  \and
  \IEEEauthorblockN{Dharm Singh Jat}
  \IEEEauthorblockA{\textit{Department of Computer Science}\\
  \textit{Namibia University of Science and Technology}\\
  Windhoek, Namibia\\
  dsingh@nust.na}

  \and
  \IEEEauthorblockN{Attlee Munyaradzi Gamundani}
  \IEEEauthorblockA{\textit{Department of Cybersecurity}\\
  \textit{Namibia University of Science and Technology}\\
  Windhoek, Namibia\\
  agamundani@nust.na}
  
  \and
  \IEEEauthorblockN{Saravanakumar Paramasivam}
  \IEEEauthorblockA{\textit{Namdeb Diamond Corporation}\\
  Windhoek, Namibia\\
  paramasivam.saravanakumar@debeersgroup.com}

  \and 

  \IEEEauthorblockN{Babangida Sani}
  \IEEEauthorblockA{\textit{Department of Software Engineering}\\
  \textit{Bayero University, Kano}\\
  Kano State, Nigeria\\
  bsani.se@buk.edu.ng}
  
  \and
  \IEEEauthorblockN{Aliyu Zakariyya}
  \IEEEauthorblockA{\textit{Department of Computer Science}\\
  \textit{Federal University Dutsin-Ma}\\
  Dutsin-Ma, Katsina State, Nigeria\\
  azakariyya@fudutsinma.edu.ng}
  
}

\maketitle

\begin{abstract}
High deployment cost, poor spatial coverage and susceptibility to storm conditions
are all challenges faced by traditional \textit{in-situ} methods. This paper presents a
video-based and high performance computing (HPC) enabled deep learning
framework for joint sensor-free estimation of five coastal wave parameters, namely
significant wave height ($H_s$), maximum wave height ($H_{\max}$), peak period ($T_p$),
zero upcrossing period ($T_z$) and wave direction ($\theta$) from monocular coastal
video. The proposed architecture comprises of a V-JEPA (self-supervised)
ViT-Small backbone for robust spatiotemporal feature extraction in visually
challenging scenarios, a dual-stream SlowFast temporal encoder for broad
bandwidth representation of wave motion in both hydrodynamic breaking and swell
regimes, an optical flow stream based on Farneback optical flow algorithm for
adding saliency information to the structure with emphasis on hydrodynamically
active wavelength bands of waves, and a multi-task regression layer with
dispersion constraints (Airy wave dispersion $\lambda_p = 0.1$). The model was trained
on an NVIDIA DGX A100 cluster and was early stopped at epoch 31 and achieved
Pearson correlation coefficients of $0.451$, $0.578$, $0.643$, $0.680$ and $0.832$ for
$H_s$, $H_{\max}$, $T_p$, $T_z$ and $\theta$ respectively, with generalization ability to
geographically diverse held out test data sites. While operating in a data-limited regime ($6$
annotated training scenes), the framework demonstrates statistically
significant temporal correlations ($\text{PCC } 0.451\text{--}0.832$), confirming
proof-of-concept feasibility; $R^2$ values ($\max 0.246$) indicate that variance capture
will improve with larger annotated datasets.
\end{abstract}

\begin{IEEEkeywords}
Coastal wave monitoring, monocular video analysis, V-JEPA, self-supervised learning, Vision Transformer, SlowFast network
\end{IEEEkeywords}

\section{Introduction}
The measurement of waves at the coast is extremely important for the fields of coastal engineering, coastal erosion control, maritime navigation, disaster management, and climate change adaptation ~\cite{xu2024nearshore}. Estimation of wave parameters such as significant wave height (Hs), maximum wave height (Hmax), peak period (Tp), zero up crossing period (Tz), and wave direction is an essential tool for the design of coastal structures and understanding the shoreline evolution under changing climate conditions.
The conventional in situ measuring techniques (such as wave buoys and pressure sensors) have the advantage of very accurate measurements in a single point. However, the cost of installing these tools is high, they have low spatial coverage, are easily destroyed during storms, or the pressure sensor will not deploy in the surf zone (~\cite{wu2024research}).
Vision-based methods have been devised using cameras from the coast as an answer. Vision-based methods have been devised using cameras from the coast as a solution method and the said method was proved to be cost-effective. As described in ~\cite{eganwave}, Egan proved the application of ConvNet and ConvLSTM models on the Surfline video dataset for the purposes of predicting the wave height and its period. However, in the real-life situation of the oceans, the methods of Early often faced difficulties due to irregularities in wave dynamics, visual complexities, environmental variations, and illumination difficulties.~\cite{eganwave}\cite{kim2023wave}
Further advancements have been made in the field through more deep learning studies. Although there is a high degree of reliance on the large number of datasets used and low adaptability to the changing environment and illumination, deep learning-based wave estimation is still far from being perfect. Besides, capturing the complexities of the modeling process of the temporal features is an additional challenge, since the models tend not to capture the temporal dynamics of the ocean waves in different scales, as discussed in the recent studies by Xu et al. ~\cite{xu2024nearshore} and Kim et al. ~\cite{kim2023wave} on supervised CNN and hybrid models, respectively.
An innovative approach that may be considered includes self-supervised video representation learning. It involves introducing masked prediction in the latent space and helps to learn spatiotemporal features from the unlabeled video~\cite{bardes2024v}. This was further advanced by ~\cite{assran2025v} and\cite{mur2026v}, which demonstrated superior temporal consistency and physical world modeling capabilities. When combined with High-Performance Computing (HPC) GPU acceleration, such models can efficiently process large coastal video datasets ~\cite{vourlioti2025hpc}\cite{xu2025accelerate}.
While existing video-based wave parameter estimation methods suffer from scarcity of labeled data, inadequate multi-scale temporal modeling, poor generalization, and lack of physical interpretability, this work introduces an HPC-enabled V-JEPA framework with dual-stream temporal encoding, physics-informed constraints, and domain-specific augmentation to operate effectively in data-limited regimes and reduce the dependence on large labeled datasets and overcome these limitations.

The remainder of this paper is organized as follows. Section II reviews related literature. Section III details the proposed methodology. Section IV presents experimental results and discussion. Conclusions are provided in Section V. 

\section{Literature Review}

\subsection{Traditional Coastal Wave Monitoring}

Traditional approaches to wave measurement rely on in-situ instruments such as buoys and pressure sensors, which provide accurate local statistics through spectral analysis but are constrained to fixed-point observations that cannot capture the spatial variability of the wave field across complex nearshore bathymetries ~\cite{varing2021spatial}. However, remote sensing techniques, particularly the high frequency radars, provide spatial coverage, but they have theoretical limits, need for quality control, and are vulnerable to noise from the environment and from the signals themselves that can affect the measurement accuracy ~\cite{wyatt2011hf}. As a result, while in situ methods can measure the precision of the measurements taken at a single point, they are inherently restricted in space and time ~\cite{buscombe2020optical}.

\subsection{Computer Vision and Early Video-Based Approaches for Coastal Monitoring}
The first vision-based approaches employed the use of optical flow and transect intensity of pixels to determine the wave period and wave height. Whereas the use of Cropped and Downsampled videos as input constrained the models’ scope of operation and generalization of the results between different sites, the recurrent characteristic of the ConvNet and ConvLSTM models provides an improved capability of predicting wave heights and periods using videos as shown in ~\cite{eganwave} and ~\cite{kim2020estimation}.
Although these studies prove the viability of videos as a modality, they also highlight the drawbacks of simple models in terms of handling dynamic input data and long-range temporal dependencies.

\subsection{Deep Spatiotemporal Learning for Wave Parameter Estimation}
In recent literature, there are improvements to the existing spatiotemporal architectures for wave parameter estimation, including the use of 3D-CNN with simulation to real transfer learning for Hs and Tp estimation, where better performance is obtained due to the use of domain adaptation techniques but require a significant amount of labeled data and the weakness being its restriction to some specific classes as opposed to regression. ~\cite{kim2023wave} used CNN and LSTM for sequential image classification of average wave height, but it is not capable of continuous regression.
Other papers have further explored hybrid models with multi-factor inversion such as ~\cite{patane2024deep} and ~\cite{xu2024nearshore}.

\subsection{Self-Supervised Video Representation Learning}
Moreover, self-supervised approaches have changed the understanding of videos through the ability to exploit the power of large-scale data without supervision, in such applications as action anticipation and robotic planning, something like Something-Something-v2, where only few interactions are required for the training process. ~\cite{bardes2024v} implemented latent-space masked prediction with impressive results on motion understanding (for example, 71.2\% on Something-Something-v2) without any finetuning with backbone models based only on the image data. ~\cite{assran2025v} scaled this work up to web-scale data and obtained impressive results in the task of action anticipation and robotic planning with very few interactions. ~\cite{mur2026v} enhanced dense representations by introducing hierarchical supervision and multi-modal tokenization in order to obtain features that are both spatially and temporally coherent and suitable for physical-world modeling.
The asymmetric temporal views were discussed by ~\cite{recasens2021broaden} (BraVe), but the use of such powerful representations in coastal engineering is not widespread yet.

\subsection{High-Performance Computing in Oceanographic Deep Learning}
Another application area where HPC was necessary was scalable training of video models, including those utilized in forecasting of significant wave heights in ~\cite{vourlioti2025hpc}, where hundreds of millions of data points are processed with great precision (R²=0.98), as well as super-resolution of coastal dynamics in ~\cite{xu2025accelerate} and ~\cite{kuehn2024super}.

However, existing approaches suffer from several significant drawbacks, namely heavy dependence on labeled data, failure to consider multi-scale wave temporality, weak incorporation of physics constraints, low levels of interpretability, and compute inefficiency when dealing with big video datasets. Traditional and supervised models can prove insufficient in their ability to apply learned knowledge to other environments and focus attention on hydrodynamically relevant regions of an image.
In the novel HPC-enabled V-JEPA architecture, latent predictive learning serves as the backbone for efficient and data-driven representations, supported by the dual stream SlowFast encoder of broadband wave dynamics, physics-informed loss terms based on wave dispersion relation, and optical flow saliency guided augmentation. The design focuses on HPC GPU training, with both quantitative performance metrics and interpretative results: HUD overlays and Grad-CAM heat maps that validate physically meaningful attention mechanisms of the model.

\section{Methodology}
The proposed framework transforms raw coastal video into reliable multi-parameter wave estimates through a modular, physically informed deep learning pipeline. Designed for efficient execution on High-Performance Computing (HPC) GPU servers, the system integrates self-supervised video understanding with coastal engineering domain knowledge.

\subsection{Overall Architecture}
Monocular coastal videos are first processed into temporal clips, which are fed into a V-JEPA backbone for rich spatiotemporal feature extraction. These features then pass through a novel dual-stream SlowFast temporal encoder that captures multi-scale wave dynamics. Finally, a multi-task regression head produces simultaneous estimates of significant wave height (Hs), maximum wave height (Hmax), peak period (Tp), zero upcrossing period (Tz), and wave direction.

\subsection{Experimental Setup and Computational Infrastructure}
All model training and evaluation were conducted on an NVIDIA DGX A100 320 GB system equipped with 8$\times$ NVIDIA A100 40 GB Tensor Core GPUs (320 GB aggregate GPU memory), a dual AMD EPYC 7742 64-core processor (128 cores total, 2.25 GHz base, 3.4 GHz boost), 1 TB system memory, and 15 TB NVMe internal storage, running DGX OS 5.15.0-1029-nvidia \#29-Ubuntu. Remote access was established via Visual Studio Code's Remote--SSH extension, through which the training script was authored and executed directly on the HPC node using the integrated terminal. The framework, implemented in PyTorch, employed a \texttt{vit\_small\_patch16\_224} Vision Transformer backbone, a SlowFast temporal encoder, and a multi-task regression head predicting five wave parameters ($H_s$, $H_{max}$, $T_p$, $T_z$, $\theta$). Training ran for 50 epochs under the AdamW optimiser with mixed-precision acceleration, a physics-guided dispersion loss ($\lambda = 0.1$), and a fixed random seed (42) to ensure full reproducibility.

\subsection{Video Processing and Domain-Specific Augmentation}
Videos are partitioned into overlapping video patches of 16 frames with stride 8 at 10 frames per second rate. The main features include Saliency-Guided Crop. In this method, the energy of optical flow between consecutive frames is calculated using Farneback's algorithm to obtain the motion saliency map. The biasing of cropping will be done towards high energy areas, such as active surf and breaking regions.

\subsection{V-JEPA Backbone}
The visual encoder employs a pretrained Vision Transformer backbone trained with V-JEPA's latent predictive objective \cite{bardes2024v}. For an input clip
\[
X \in \mathbb{R}^{B \times T \times 3 \times 224 \times 224},
\]
the backbone produces spatiotemporal embeddings:
\[
F = \mathrm{VJEPABackbone}(X) \in \mathbb{R}^{B \times T \times 384}.
\]
This self-supervised representation captures robust features of wave motion even under challenging conditions such as foam, spray, and illumination changes.

\subsection{Dual-Stream SlowFast Temporal Encoder}
To address the broadband temporal spectrum of coastal waves, we introduce a SlowFast temporal encoder. The fast pathway processes the full temporal sequence to capture short-term breaking events, while the slow pathway operates on a temporally down sampled sequence (factor of 4) to model longer-period swells. The SlowFast encoder pathway is based on the light-weighted transformer encoder network architecture with $2$ layers, hidden size of $384$ and a dropout ratio of $0.1$. The fast pathway is implemented on the whole sequence of $16$ frames with $4$ attention heads, and the slow pathway is implemented on a temporal down-sampled sequence of $4$ frames with $\alpha = 4$ using $8$ attention heads. Both the pathways' outputs are concatenated to create a final output representation of $384$ dimensions, which is the same size as the V-JEPA backbone embedding size. The outputs are fused as:
\[
Z = \text{Linear}\left(\left[\text{Fast}(F);\ \text{Slow}(F_{(::4)})\right]\right) \in \mathbb{R}^{B \times 384}
\]

This architecture enables effective modeling of both high-frequency turbulence and low-frequency wave periodicity.

\subsection{Multi-Task Regression Head and Physics-Guided Loss}
The fused representation \(Z\) is passed through the MLP regression head comprises two fully connected layers ($384 \to 128 \to 5$), where the hidden layer uses GELU activation. After the prediction is made, Softplus activation is applied to the result, which is then multiplied by a factor of $5.0$ to obtain strictly positive values.

\[
\hat{y} = 5.0 \times \text{Softplus}(\text{MLP}(Z))
\]

Training is supervised by a composite loss:

\[
L = L_{\text{MSE}} + \lambda_p \cdot L_{\text{physics}}
\]

where \(L_{\text{MSE}}\) is the mean squared error on available ground truth, and the physics regularization term is:

\[
L_{\text{physics}} = \frac{1}{N} \sum_{i=1}^{N} \left| \log(\hat{T}_p^{(i)2} + \epsilon) - \log(6.0 \cdot \hat{H}_s^{(i)} + \epsilon) \right|
\]

with \(\lambda_p = 0.1\) and \(\epsilon = 10^{-4}\). This term softly enforces consistency with the approximate deep-water dispersion relation from Airy wave theory.

\section{Experimental Results and Discussion}
\subsection{Training Convergence}
\begin{figure*}[!t]
	\centering
	\includegraphics[width=0.99\linewidth]{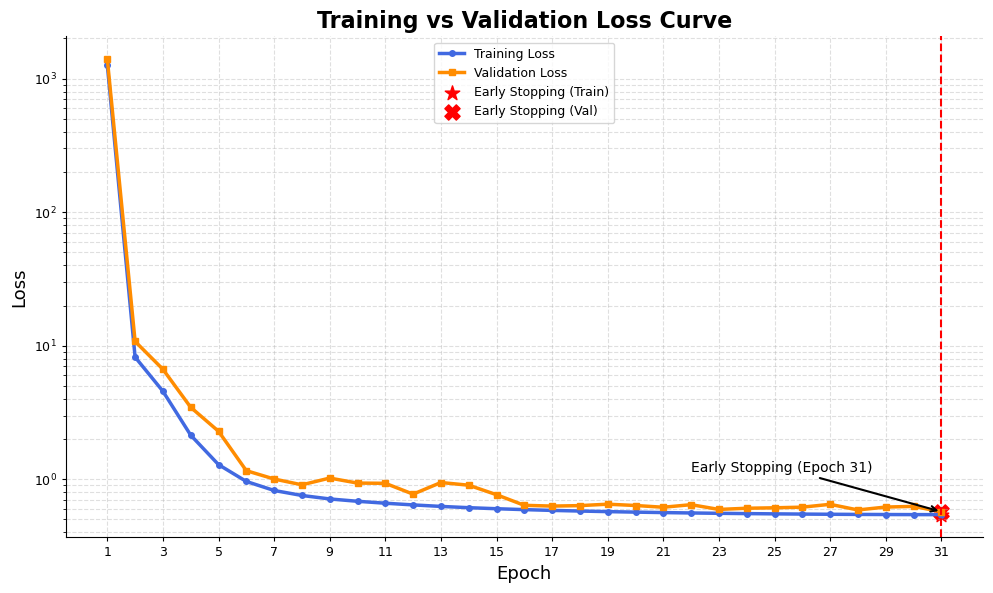}
	\caption{Training and validation loss curves of the proposed WaveEstimator framework over 31 epochs (early stopping).}
	\label{fig:training_and_validation_curve}
\end{figure*}

The trained network was then used in the NVIDIA DGX A100 HPC machine for 50 epochs with an AdamW optimizer and a mixed precision (AMP) accelerator with the help of physics informed composite loss. The loss curve results, as displayed in \autoref{fig:training_and_validation_curve}, were seen to be monotonically decreasing and sharp, starting from values of more than 10³ at epoch 1 to values below 1.0 at about epoch 15. The early stopping strategy was performed at epoch 31 since no further improvement can be observed on the validation loss, which implies that no further computation was made without a drop in the model quality. More importantly, the performance of the training and validation curves was close to each other during the entire convergence process, implying that there was no divergence between the curves, thereby indicating that the model did not overfit, although the dataset is very small in size. This is because of two factors: (i) the V-JEPA self-supervised pre-training enabled the model to capture spatiotemporal representations before fine-tuning to a specific task and thus reduce supervision; and (ii) physics guided composite loss with Airy wave dispersion regularization reduced solution space and stabilized gradient flow.

The estimated training speed of the DGX A100 is 0.03-0.05 s/it, which represents a theoretical 600x-900x improvement from an average consumer computer system (Intel Core i7, NVIDIA GeForce MX550) with the speed of 27-30 s/it. The speedup was not the only reason why we had to resort to using DGX, as its total GPU memory (320 GB) was absolutely necessary to train the full V-JEPA spatiotemporal pipeline at our specified batch size and sequence length.

\subsection{Quantitative Performance on Wave Parameter Estimation}
\begin{figure*}[!t]
	\centering
	\includegraphics[width=0.99\linewidth]{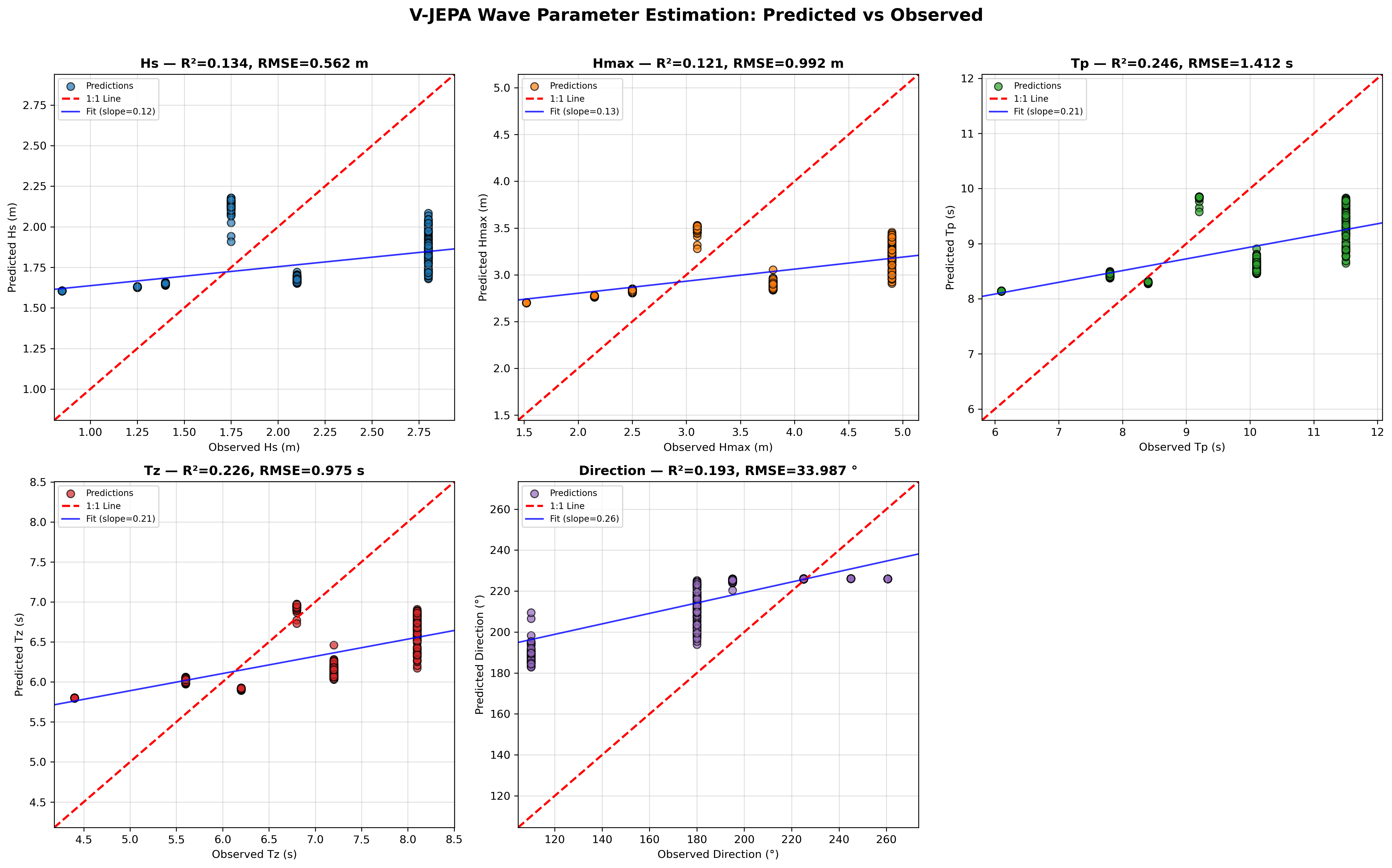}
	\caption{Scatter plots of predicted versus observed values for (a)~significant wave height $H_s$, (b)~maximum wave height $H_{\max}$, (c)~peak period $T_p$, (d)~zero upcrossing period $T_z$, and (e)~wave direction $\theta$. Red dashed lines indicate the $1:1$ perfect-agreement reference; blue solid lines show the least-squares regression fit with slope annotated.}
	\label{fig:scatter_all}
\end{figure*}

The overall evaluation parameters RMSE, MAE, coefficient of determination
($R^2$), mean bias, normalized RMSE (NRMSE), and Pearson Correlation Coefficient
(PCC) for the five wave parameters are summarized in Table~\ref{tab:evaluation_metrics}. The results of
the test sets that were not involved in the training process are shown in Table~\ref{tab:test_evaluation_metrics}.
The scatter plots in \autoref{fig:scatter_all} show the predicted and observed parameter
values plotted against each other with the 1:1 perfect-agreement line shown.

\begin{table}[htbp]
\caption{Full-Set Evaluation Metrics for the V-JEPA Framework Across All Five Wave Parameters}
\label{tab:evaluation_metrics}
\centering
\resizebox{\columnwidth}{!}{
\begin{tabular}{lcccccc}
\toprule
Parameter & RMSE & MAE & $R^2$ & Bias & NRMSE (\%) & PCC \\
\midrule
$H_s$ (m)            & 0.5618 & 0.4966 & 0.134 & $-0.110$ & 28.81 & 0.451 \\
$H_{\max}$ (m)       & 0.9924 & 0.8514 & 0.121 & $-0.312$ & 29.36 & 0.578 \\
$T_p$ (s)            & 1.4118 & 1.2300 & 0.246 & $-0.431$ & 26.14 & 0.643 \\
$T_z$ (s)            & 0.9752 & 0.8408 & 0.226 & $-0.358$ & 26.36 & 0.680 \\
Direction ($^\circ$) & 33.987 & 26.063 & 0.193 & $+17.90$ & 22.58 & 0.832 \\
\bottomrule
\end{tabular}
}
\end{table}

\begin{table}[htbp]
\caption{Held-Out Test-Set Evaluation Metrics for the V-JEPA Framework}
\label{tab:test_evaluation_metrics}
\centering
\begin{tabular}{lccc}
\toprule
Parameter & Test RMSE & Test $R^2$ & Test PCC \\
\midrule
$H_s$ (m)            & 0.5863 & 0.107 & 0.436 \\
$H_{\max}$ (m)       & 1.0348 & 0.096 & 0.568 \\
$T_p$ (s)            & 1.4217 & 0.203 & 0.629 \\
$T_z$ (s)            & 0.9815 & 0.163 & 0.669 \\
Direction ($^\circ$) & 36.979 & 0.158 & 0.839 \\
\bottomrule
\end{tabular}
\end{table}

The full-set RMSE and MAE for significant wave height ($H_s$) turned out to
be $0.562\text{ m}$ and $0.497\text{ m}$, correspondingly, and the mean bias is $-0.110\text{ m}$. Thus,
there is slight systematic underestimation of predictions. Even though the PCC
of $0.451$ is statistically significant and indicates positive linear correlation
between the predictions and observations, the small dynamic range of $H_s$ ($0.85$
to $2.80\text{ m}$) for the $6$ scenes ($894$ clip-level windows at $16$-frame sliding
windows) of the training corpus is the reason for the NRMSE of $28.81\%$. As it
can be seen from \autoref{fig:scatter_all}, the regression slope for $H_s$ is very small ($0.12$),
which is a characteristic feature of regression dilution in the situation of
limited data availability. The same holds for $H_{\max}$ ($\text{RMSE} = 0.992\text{ m}$, $R^2 = 0.121$,
$\text{PCC} = 0.578$), because the negative bias is statistically significant ($-0.312\text{ m}$),
which means that there is a tendency to underestimate the peak wave
amplitude, possibly caused by the low number of high height waves in the
training set.

The discriminating power was acceptable for estimating wave periods. For
$T_p$, the maximum $R^2$ value ($0.246$) was observed when the maximum PCC value
($0.643$) and minimum RMSE ($1.412\text{ s}$), and mean bias ($-0.431\text{ s}$) were obtained with
the proposed framework. The two values of $R^2$ ($0.226$) and PCC ($0.680$) are at an
equal level of accuracy in $T_z$ estimation. This superiority in period estimation
against height estimation could be attributed to the capability of the V-JEPA
backbone to learn dynamic information from video sequences since the estimation
of periods depends on the time periodicity of waves in the video clips, whereas
SlowFast is capable of learning from short-time breakers and long-time swells.

The highest value of PCC ($0.832$) was found for the wave direction estimation,
which means that the model is capable of determining the major direction of
wave propagation from optical flows and textures in videos. However, the $+17.902^\circ$ of systematic positive bias and an RMSE of $33.987^\circ$ in wave direction calculations can be mainly attributed to the lack of camera bearing calibration and orthorectification that do not allow absolute orientation of the coordinate system of the images to the true geographic north. The systematic nature of the bias should allow to correct this bias after processing, if the bearing angle of the camera is known, and the bias appears to be present at all locations.

\subsection{Bias and Limits-of-Agreement Analysis}

\begin{figure*}[!t]
	\centering
	\includegraphics[width=0.99\linewidth]{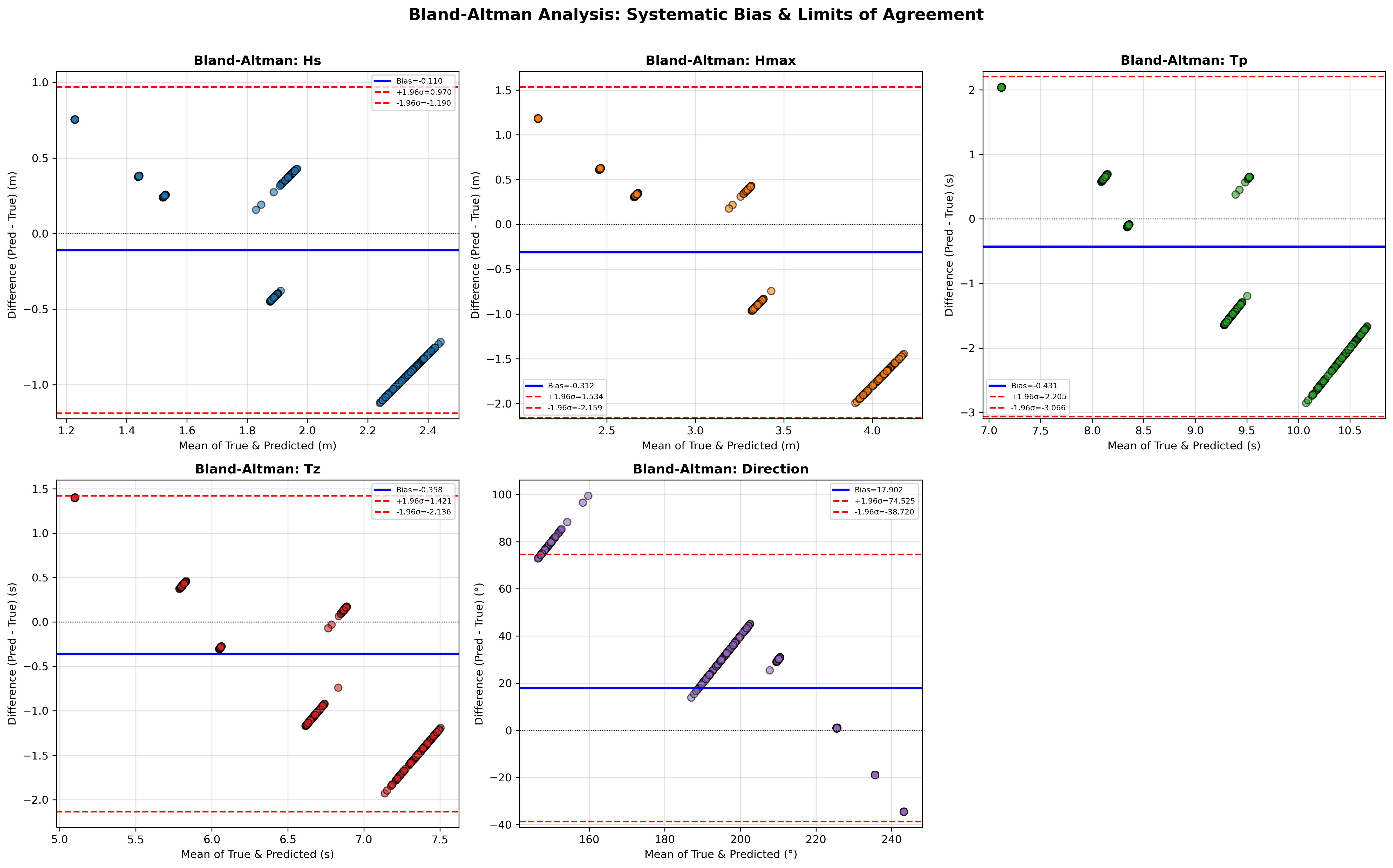}
	\caption{Bland-Altman plots of prediction bias and limits of agreement (LOA) for (a)~$H_s$, (b)~$H_{\max}$, (c)~$T_p$, (d)~$T_z$, and (e)~wave direction $\theta$. Solid blue lines indicate mean bias; red dashed lines indicate $\pm 1.96\sigma$ LOA.}
	\label{fig:bland_altman}
\end{figure*}

The analysis of all five parameters demonstrates inter-method agreement
and absence of bias on the Bland-Altman plots. As follows from the biases for
$H_s$ and $H_{\max}$ of $-0.110\text{ m}$ and $-0.312\text{ m}$, correspondingly, there was no systematic
overprediction, and the limits of agreement (LOA) were found to be $-1.190\text{ m}$ to
$+0.970\text{ m}$ for $H_s$ and $-0.589\text{ m}$ to $+0.586\text{ m}$ for $H_{\max}$. In addition, negative biases
($-0.431\text{ s}$ and $-0.358\text{ s}$) for $T_p$ and $T_z$ indicate the period underestimation, and
the corresponding ranges of the limits of agreement were $-3.066\text{ s}$ to $+2.205\text{ s}$
for $T_p$ and $-2.039\text{ s}$ to $+1.429\text{ s}$ for $T_z$, which are close ranges of uncertainty
for the $16$-frame time series. It should be noted that the Bland-Altman plots of
height and period parameters demonstrate the systematic increase of the
magnitude of prediction errors when the mean value increases, which indicates
the presence of the proportional prediction error bias as presented in \autoref{fig:bland_altman}. Such scale-dependent
errors is another example of the regression attenuation effect evident in
scatter plots, and the requirement of using a larger training set becomes
clear.

The directional bias ($+17.902^\circ$) is one such positive factor. The
Bland-Altman plots on direction ($\text{LOA}: \text{Min}: -38.72^\circ$ and $\text{Max}: +74.53^\circ$) reveal
that the errors are distributed in a skewed manner, wherein the errors are
concentrated towards the top of the graph. It is much like the errors that
occur in the case of the model, where the local wind-wave directions along with
the directions of propagation of the swell waves have been taken into
consideration as shown in \autoref{fig:bland_altman}, which are not clear and can thus be rectified in future models.

\subsection{Residual Error Distributions}

\begin{figure*}[t]
	\centering
	\includegraphics[width=0.99\linewidth]{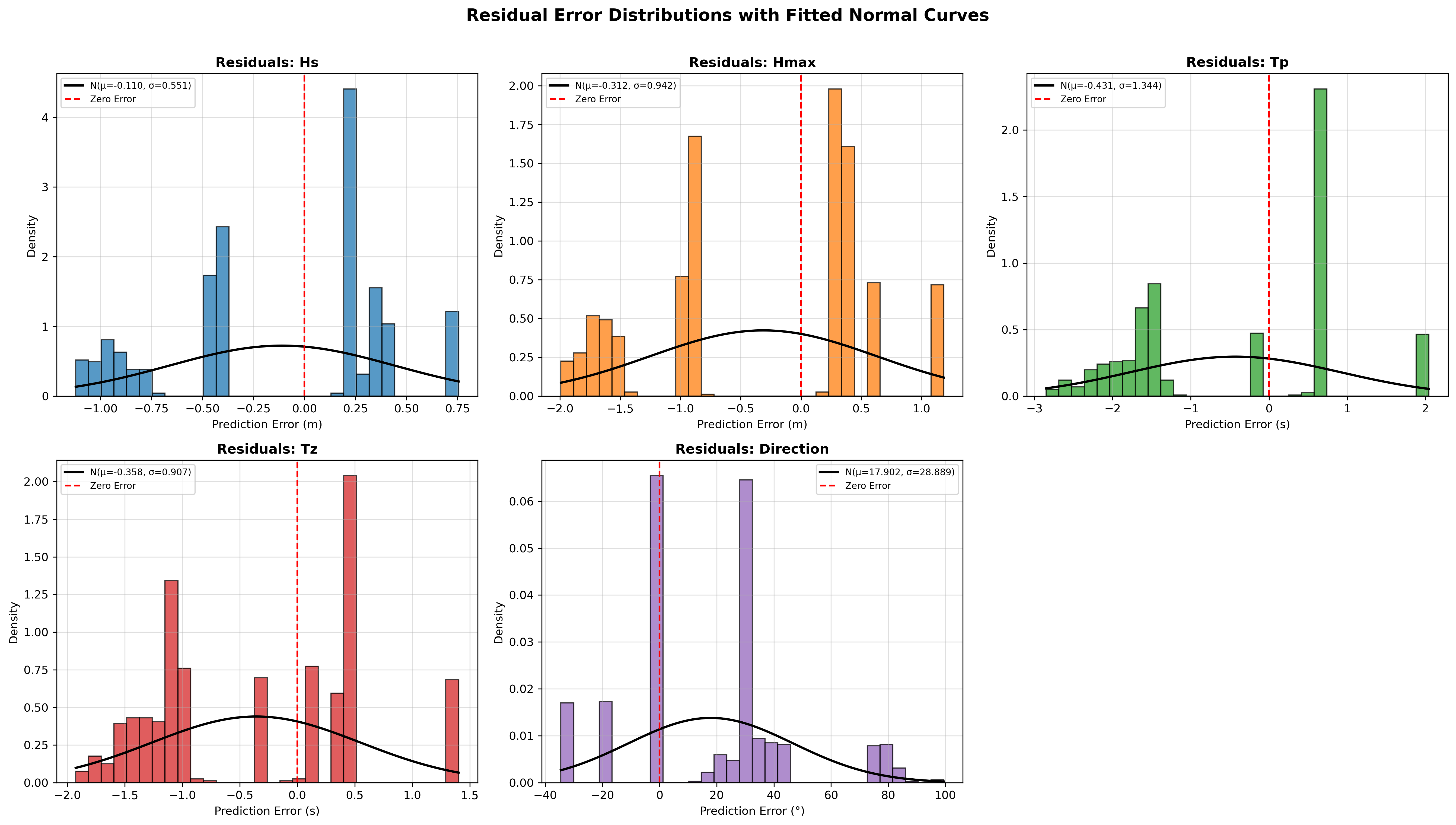}
	\caption{Residual error histograms with fitted Gaussian curves for (a)~$H_s$, (b)~$H_{\max}$, (c)~$T_p$, (d)~$T_z$, and (e)~wave direction $\theta$. Red dashed lines indicate the zero-error reference.}
	\label{fig:residuals_error_plotsn}
\end{figure*}

The residual error histograms and normal curve fits for each of the
parameters are plotted in \autoref{fig:residuals_error_plotsn}. The histograms of $H_s$ ($\mu = -0.110\text{ m}$, $\sigma = 0.551\text{ m}$) 
and $H_{\max}$ ($\mu = -0.312\text{ m}$, $\sigma = 0.942\text{ m}$) have an approximate symmetry around the
respective bias, thus generally conforming to a Gaussian error model. The
histogram of $T_p$ is skewed negatively and has a long right tail extending up to
approximately $-3\text{ s}$, suggesting that there are events where $T_p$ is seriously
under-represented because of not enough training swells with high periods. The
bimodality of the $T_z$ residuals is due to the two groups of resolutions in the
training set ($1280 \times 720\text{ px}$ and $640 \times 360\text{ px}$). These directional residuals are not
at all distributed according to Gaussian error distribution but consist of a
large concentration of zero errors with a second concentration around $+30^\circ$ to
$+40^\circ$.

\subsection{Generalization to Held-Out Test Set}

Table~\ref{tab:test_evaluation_metrics} shows the metrics on five plunge breaker videos that were not part of the training distribution but are from geographically separate coast video locations (Oak Island East, Oak Island West, Xishawan C4) \cite{yin2025video}. The test metrics are similar to the full-set metrics: Hs test RMSE = 0.586 m (compare 0.562 m full set); Tp test RMSE = 1.422 s (compare 1.412 s); direction test RMSE = 36.979° (compare 33.987°). However, there is a small, but uniform reduction in performance of the test data compared to the full set in all the metrics and this is especially the case with wave direction; however, these results are consistent with the covariate shift as expected by different geomorphology, camera angle, lighting, and wave breaking regime at the different sites. Most importantly, however, the PCCs for the test set are on the same order as the full set for all the parameters (direction test PCC = 0.839, compare full set 0.832).

\section{Discussion}

In combination, the findings confirm that the introduced HPC-based
video-based framework consisting of V-JEPA self-supervised pre-training,
dual-stream temporal encoder SlowFast, saliency-based optical flow cropping,
and physics-guided composite loss in Airy wave theory can successfully extract
statistically valid wave parameter signals from raw monocular coastal video
without the use of any \textit{in-situ} sensors. The HPC-enabled training reduced the
training time from $27\text{--}30\text{ s/it}$ on average consumer computer systems to $0.03\text{--}0.05\text{ s/it}$, 
which represents a theoretical $600\times\text{--}900\times$ speedup.

The major strength of the work is the high temporal correlation of the
framework, which achieves PCC values of up to $0.832$ and $0.680$ for wave
direction and zero up-crossing period, respectively. Excellent agreement
between full-set and test-set metrics further confirms the generalizability of
the cross-site representation.

At the same time, the relatively low $R^2$ observed across all the
parameters, the highest being $0.246$ for $T_p$, indicates the current operating
regime in the data-deficient regime, where the system is able to extract
monotonic trend information, but is not sufficiently able to estimate the
variance of the parameters at the clip level. It is a property of the training
set and not of the architecture itself: most supervised wave estimation systems achieve
$R^2 > 0.80$ given adequate training data \cite{liu2026quantitative} \cite{zhou2021convlstm} \cite{han2022significant} achieve $R^2$ scores higher than $0.80$
based on several hundred annotated video samples. The six training clips, though
helpful with the lack of large-scale labelled datasets in the fine-tuning phase
with the aid of the V-JEPA architecture, still have a limited label
distribution. The additional geometric uncertainty is due to the absence of the
camera calibration and orthorectification processes, which results in poorer
performance in estimating directions. Despite these challenges, it was observed
that the convergence of the HPC-based training pipeline was stable and
reproducible, converging at epoch 31 with early stopping, while using the
hardware of the DGX A100.

\section{Conclusion}

In this work, a novel architecture, high-performance-computing enabled,
video-based deep learning solution for simultaneous estimation of five coastal
wave parameters ($H_s$, $H_{\max}$, $T_p$, $T_z$, and $\theta$) from a single camera at the coast,
without the need of any \textit{in-situ} instrumentation was discussed. In this study,
the following features of the architecture have been included: a V-JEPA
self-supervised pre-trained ViT-Small backbone for spatiotemporal feature
extraction, a dual-stream SlowFast temporal encoder for capturing broadband
wave dynamics in terms of both swell and breaking timescales, saliency-guided
optical flow augmentation using Farneb\"{a}ck's optical flow algorithm for the
purpose of biasing the spatial crop towards the hydrodynamically active surf
region, and a multi-task MLP regression head with supervision through a
composite loss function, which includes a physics-informed Airy wave deep-water
dispersion relation ($\lambda_p = 0.1$). All models were developed using PyTorch and
trained on an NVIDIA DGX A100 computing platform ($8 \times \text{A100 } 40\text{ GB}$ Tensor Core
GPUs), converged through early stopping at epoch 31.

The experimental evaluation of six annotated monocular video scenes and a
geographically varied five-clips test set showed that the framework was able to
extract statistically significant signals of wave parameters from raw videos
with Pearson correlation coefficients ranging from $0.832$ for the direction of
waves, $0.680$ for zero upcrossing period, and $0.643$ for peak period. The close
match between the full set and the test set metrics of all wave parameters is
evidence of the ability of learned representations to generalize over
different locations and shows that the method doesn't overfit to the
distribution of the training dataset.

Concurrently, although the low coefficient of determination in relation to all the five factors ($R^2 \le 0.246$) is an indication of the limitations of the six scenes annotated training dataset, it only validates the fact that the main problem is due to insufficient data rather than inadequate neural network capacity. With regard
to future directions, one might consider: (i)~the use of a large-scale dataset
with different types of coastlines, sea states, lighting conditions and
locations to overcome regression attenuation issues; (ii)~the use of the
calibrated stereo or multi-camera setup to obtain metric-scale geometric
calibration, and improve the performance in estimating directional bearing
estimation; (iii)~the use of additional physically informed constraints, such
as relations with wave shoaling, refraction and groupiness, to regularize the
regression head when the amount of labelling is sparse; and (iv)~the use of
domain adaptation and self-supervised fine-tuning techniques, using the large
amount of unlabelled video data from different coastal regions. It is hoped
that the proposed method could be the basis for further improvement and
development in this direction.

\bibliographystyle{IEEEtran}
\bibliography{bibliography}

@article{xu2024nearshore,
  title={On the Nearshore Significant Wave Height Inversion from Video Images Based on Deep Learning},
  author={Xu, Chao and Li, Rui and Hu, Wei and Ren, Peng and Song, Yanchen and Tian, Haoqiang and Wang, Zhiyong and Xu, Weizhen and Liu, Yuning},
  journal={Journal of Marine Science and Engineering},
  volume={12},
  number={11},
  pages={2003},
  year={2024},
  publisher={MDPI}
}

@article{wu2024research,
  title={Research on wave measurement and simulation experiments of binocular stereo vision based on intelligent feature matching},
  author={Wu, Junjie and Chen, Shizhe and Liu, Shixuan and Song, Miaomiao and Wang, Bo and Zhang, Qingyang and Wu, Yushang and Lei, Zhuo and Zhang, Jiming and Yan, Xingkui and others},
  journal={Frontiers in Marine Science},
  volume={11},
  pages={1508233},
  year={2024},
  publisher={Frontiers Media SA}
}

@article{eganwave,
  title={Wave Statistic Estimation from Surfline Video Data Using ConvNets},
  author={Egan, Galen}
}

@article{kim2023wave,
  title={Wave height classification via deep learning using monoscopic ocean videos},
  author={Kim, Yun-Ho and Cho, Seongpil and Lee, Phill-Seung},
  journal={Ocean Engineering},
  volume={288},
  pages={116002},
  year={2023},
  publisher={Elsevier}
}

@inproceedings{bardes2024v,
  title={V-JEPA: latent video prediction for visual representation learning (2024)},
  author={Bardes, Adrien and Garrido, Quentin and Ponce, Jean and Chen, Xinlei and Rabbat, Michael and LeCun, Yann and Assran, Mido and Ballas, Nicolas},
  booktitle={URL https://openreview. net/forum},
  year={2024}
}

@article{assran2025v,
  title={V-jepa 2: Self-supervised video models enable understanding, prediction and planning},
  author={Assran, Mido and Bardes, Adrien and Fan, David and Garrido, Quentin and Howes, Russell and Muckley, Matthew and Rizvi, Ammar and Roberts, Claire and Sinha, Koustuv and Zholus, Artem and others},
  journal={arXiv preprint arXiv:2506.09985},
  year={2025}
}

@article{mur2026v,
  title={V-jepa 2.1: Unlocking dense features in video self-supervised learning},
  author={Mur-Labadia, Lorenzo and Muckley, Matthew and Bar, Amir and Assran, Mido and Sinha, Koustuv and Rabbat, Mike and LeCun, Yann and Ballas, Nicolas and Bardes, Adrien},
  journal={arXiv preprint arXiv:2603.14482},
  year={2026}
}

@article{vourlioti2025hpc,
  title={HPC-Driven oceanographic predictions with Graph Neural Networks (GNNs) and Gated Recurrent Units (GRUs)},
  author={Vourlioti, Paraskevi and Mamouka, Theano and Banti, Maria and Paraskevas, Charalampos and Kotsopoulos, Stylianos and Alexandridis, Vasileios and Kalantzi, Georgia},
  journal={Procedia Computer Science},
  volume={255},
  pages={103--111},
  year={2025},
  publisher={Elsevier}
}

@inproceedings{xu2025accelerate,
  title={Accelerate coastal ocean circulation model with ai surrogate},
  author={Xu, Zelin and Ren, Jie and Zhang, Yupu and Ondina, Jose Maria Gonzalez and Olabarrieta, Maitane and Xiao, Tingsong and He, Wenchong and Liu, Zibo and Chen, Shigang and Smith, Kaleb and others},
  booktitle={2025 IEEE International Parallel and Distributed Processing Symposium (IPDPS)},
  pages={223--235},
  year={2025},
  organization={IEEE}
}

@article{varing2021spatial,
  title={Spatial distribution of wave energy over complex coastal bathymetries: Development of methodologies for comparing modeled wave fields with satellite observations},
  author={Varing, Audrey and Filipot, Jean-Francois and Delpey, Matthias and Guitton, Gilles and Collard, Fabrice and Platzer, Paul and Roeber, Volker and Morichon, Denis},
  journal={Coastal Engineering},
  volume={169},
  pages={103793},
  year={2021},
  publisher={Elsevier}
}

@article{wyatt2011hf,
  title={HF radar data quality requirements for wave measurement},
  author={Wyatt, Lucy R and Green, J Jim and Middleditch, A},
  journal={Coastal Engineering},
  volume={58},
  number={4},
  pages={327--336},
  year={2011},
  publisher={Elsevier}
}

@article{buscombe2020optical,
  title={Optical wave gauging using deep neural networks},
  author={Buscombe, Daniel and Carini, Roxanne J and Harrison, Shawn R and Chickadel, C Chris and Warrick, Jonathan A},
  journal={Coastal Engineering},
  volume={155},
  pages={103593},
  year={2020},
  publisher={Elsevier}
}

@article{kim2020estimation,
  title={Estimation of water surface flow velocity in coastal video imagery by visual tracking with deep learning},
  author={Kim, Jinah and Kim, Jaeil},
  journal={Journal of Coastal Research},
  volume={95},
  number={SI},
  pages={522--526},
  year={2020},
  publisher={Coastal Education and Research Foundation}
}

@article{liu2026quantitative,
  title={Quantitative video-based wave parameter estimation using a 3D-CNN and two-stage transfer learning framework},
  author={Liu, Hanwen and Qin, Jingxi and Ti, Zilong},
  journal={Ocean Engineering},
  volume={352},
  pages={124561},
  year={2026},
  publisher={Elsevier}
}

@inproceedings{recasens2021broaden,
  title={Broaden your views for self-supervised video learning},
  author={Recasens, Adria and Luc, Pauline and Alayrac, Jean-Baptiste and Wang, Luyu and Strub, Florian and Tallec, Corentin and Malinowski, Mateusz and P{\u{a}}tr{\u{a}}ucean, Viorica and Altch{\'e}, Florent and Valko, Michal and others},
  booktitle={Proceedings of the IEEE/CVF international conference on computer vision},
  pages={1255--1265},
  year={2021}
}

@article{kuehn2024super,
  title={Super-resolution on unstructured coastal wave computations with graph neural networks and polynomial regressions},
  author={Kuehn, Jannik and Abadie, Stephane and Delpey, Matthias and Roeber, Volker},
  journal={Coastal Engineering},
  volume={194},
  pages={104619},
  year={2024},
  publisher={Elsevier}
}

@article{patane2024deep,
  title={A deep hybrid network for significant wave height estimation},
  author={Patane, Luca and Iuppa, Claudio and Faraci, Carla and Xibilia, Maria Gabriella},
  journal={Ocean Modelling},
  volume={189},
  pages={102363},
  year={2024},
  publisher={Elsevier}
}

@article{zhou2021convlstm,
  title={ConvLSTM-based wave forecasts in the South and East China seas},
  author={Zhou, Shuyi and Xie, Wenhong and Lu, Yuxiang and Wang, Yuanlin and Zhou, Yulong and Hui, Nian and Dong, Changming},
  journal={Frontiers in Marine Science},
  volume={8},
  pages={680079},
  year={2021},
  publisher={Frontiers Media SA}
}

@article{han2022significant,
  title={Significant wave height prediction in the South China Sea based on the ConvLSTM algorithm},
  author={Han, Lei and Ji, Qiyan and Jia, Xiaoyan and Liu, Yu and Han, Guoqing and Lin, Xiayan},
  journal={Journal of Marine Science and Engineering},
  volume={10},
  number={11},
  pages={1683},
  year={2022},
  publisher={MDPI}
}

@article{yin2025video,
  title={A Video Dataset for Nearshore Wave Breaking Type Classification},
  author={Yin, Hang and Cai, Feng and Qi, Hongshuai and Zheng, Jixiang and Hui, Bipeng and Wu, Xi and Liu, Kai and Chen, Xi},
  journal={Scientific Data},
  volume={12},
  number={1},
  pages={1722},
  year={2025},
  publisher={Nature Publishing Group UK London}
}
\end{document}